\documentclass[reprint,
amsmath,
amssymb,
aps,
pra,
]{revtex4-2}

\usepackage{booktabs} 
\usepackage{mathrsfs} 

\newcommand{\dbar}{\bar{\partial}}
\newcommand{\dif}{\mathrm{d}}
\newcommand{\p}{\partial}

\newcommand{\be}{\begin{equation}}
\newcommand{\ee}{\end{equation}}

\newcommand{\bea}{\begin{equation}\begin{aligned}}
\newcommand{\eea}{\end{aligned}\end{equation}}

\newcommand{\half}{\tfrac{1}{2}}

\newcommand{\op}{\operatorname}

\newcommand{\la}{\langle}
\newcommand{\ra}{\rangle}

\newcommand{\C}{\mathbb{C}}
\newcommand{\R}{\mathbb{R}}
\newcommand{\Z}{\mathbb{Z}}

\newcommand{\CP}{\mathbb{CP}}
\newcommand{\PT}{\mathbb{PT}}

\newcommand{\mc}{\mathcal}

\newcommand{\im}{\mrm{i}}

\newcommand{\cM}{\mathcal{M}}
\newcommand{\cN}{\mathcal{N}}

\newcommand{\cPT}{\mathcal{PT}}

\newcommand{\mf}{\mathfrak}

\newcommand{\fsl}{\mathfrak{sl}}

\newcommand{\mrm}[1]{\mathrm{#1}}

\renewcommand{\cha}{\mathrm{ch}}
\newcommand{\chern}{\mathrm{c}}

\newcommand{\Cha}{\mathrm{Ch}}

\newcommand{\Todd}{\mathrm{Td}}

\newcommand{\gSU}{\mathrm{SU}}

\newcommand{\mscr}{\mathscr}
\newcommand{\Oo}{\mathscr{O}}

\begin{document}

\title{The One-Loop QCD $\beta$-Function as an Index}

\author{Roland Bittleston}
\author{Kevin Costello}
\affiliation{Perimeter Institute for Theoretical Physics,\\ 31 Caroline Street, Waterloo, Ontario, Canada}

\begin{abstract}
In this letter we show that the one-loop QCD $\beta$-function can be obtained from an index theorem on twistor space. This is achieved by recalling that the $\theta$-angle of self-dual gauge theory flows according the one-loop $\beta$-function. Rewriting self-dual gauge theory as a holomorphic theory on twistor space this flow can be computed as the anomaly to scale invariance. The one-loop Weyl anomaly coefficient $a-c$ can be recovered similarly.
\end{abstract}

\maketitle


\section{Introduction}

The computation of the one-loop QCD $\beta$-function \cite{Politzer:1973fx,Gross:1973ju} was a milestone in the study of quantum field theory. The initial computation was quite difficult, although  it has since been streamlined by background field methods \cite{Abbott:1980hw,Abbott:1981ke}. In this Letter we give an entirely new and very simple derivation which uses only algebro-geometric methods, namely twistor theory and the index theorem.  We also give an index-theoretic computation of the one-loop Weyl anomaly coefficient $a-c$.  The numerical factors that appear in the $\beta$-function and Weyl anomaly emerge from the Bernoulli numbers in the index theorem.


\section{$\theta$-Angles in SDYM and QCD} \label{sec:theta_angle}

Let us review the self-dual limit of Yang-Mills theory \cite{Chalmers:1996rq,Losev:2017qrj}.  The fields of self-dual Yang-Mills include a gauge field $A$ and an adjoint-valued anti-self-dual $2$-form $B$. One can also couple to matter fields, but for now we will focus on the pure gauge theory.  The action is
\be \label{eqn:sdYM}
\int_{\R^4}\op{tr}\big(B \wedge F(A)_-\big) + \frac{\im}{4\pi}\tau\int_{\R^4}\op{tr}\big(F(A) \wedge F(A)\big)
\ee
where $\tau$ is a complex parameter. We can view $\theta_{SD} = - 2\pi\im\tau$ as the $\theta$-angle of the self-dual theory. We will see momentarily that, upon deforming to full Yang-Mills theory, it differs in a crucial way from the standard $\theta$-angle. We work in Euclidean signature so that instantons appear as solutions to the classical equations of motion. In order for their contributions to be suppressed in the path integral we require that $\op{Im}\tau\geq0$. In the usual way $\tau\sim\tau+1$ on account of the integrality of the Chern character. 

Perturbative QCD can be recovered as a deformation of self-dual QCD by adding the term $- \tfrac{g^2_\text{YM}}{8}\op{tr}(B^2)$ to the Lagrangian.  Integrating out $B$ the action becomes
\be \label{eqn:lagrangian_integrated}
\frac{2}{g^2_\text{YM}}\int_{\R^4}\op{tr}\big(F(A)_-\wedge F(A)_-\big) + \frac{\im}{4\pi}\tau\int_{\R^4}\op{tr}\big(F(A)\wedge F(A)\big)\,. \ee
Since $F(A)_- = \half ( F(A) - \ast F(A))$, 
\bea
&\op{tr}\big(F(A)_-\wedge F(A)_-\big) = \frac{1}{2}\op{tr}\big(F(A)\wedge F(A)\big) \\
&- \frac{1}{2}\op{tr}\big(F(A)\wedge\ast F(A)\big)\,,
\eea
and the action \eqref{eqn:lagrangian_integrated} is equivalent to
\bea
&- \frac{1}{g^2_\text{YM}} \int_{\R^4} \op{tr}\big(F(A) \wedge \ast F(A)\big) \\
&+ \bigg(\frac{1}{g_\text{YM}^2} + \frac{\im}{4\pi}\tau\bigg)\int_{\R^4} \op{tr}\big(F(A)\wedge F(A)\big)\,.
\eea
(We are using the same conventions as \cite{Schafer:1996wv}, where the Yang-Mills action evaluates to $8\pi^2k/g_\text{YM}^2$ on an instanton of charge $k$.)
 
Writing $\theta_{QCD}$ for the QCD $\theta$-angle in Euclidean signature, we have the relation
\be
\tau = \frac{\im}{2\pi}\theta_{SD} = \frac{\im}{2\pi}\theta_{QCD} + \frac{4\pi\im}{g_\text{YM}^2}\,.
\ee
Recalling that the Euclidean $\theta$-angle differs from its Lorentzian counterpart by a factor of $\im$, $\tau$ can be identified with the complex coupling of QCD.

We know that, to leading order in $\theta_{QCD}$, parity symmetry prevents $\theta_{QCD}$ from flowing under RG.  Since $g_\text{YM}$ definitely does flow, this tells us that $\theta_{SD}$ must also flow \footnote{The flow of $\theta_{SD}$ was computed by a direct Feynman diagram analysis  in \cite{Losev:2017qrj}.} , and in such a way so as to leave $\theta_{QCD}$ unchanged.  

More precisely, the variation of $\theta_{SD}$ with respect to the logarithm of the scale is
\be
\frac{\p \theta_{SD}} {\p \log \mu} = - \frac{16\pi^2}{g_\text{YM}^3}\frac{\p g_\text{YM}}{\p \log \mu}
\ee
Recall the standard formula for the one-loop $\beta$-function
\be
\frac{\p g_\text{YM}}{\p \log \mu} = - \frac{g_\text{YM}^3}{16 \pi^2} b
\ee
where
\be \label{eqn:beta1}
b = \frac{11}{3} N_c - \frac{2}{3} N_f - \frac{1}{6} N_s\,.
\ee
Here $N_f,N_s$ are the number of fundamental Dirac fermions and fundamental complex scalars respectively.  The factors of $16\pi^2/g^3_\text{YM}$ cancel leaving
\be \label{eqn:thetaSD_flow} \frac{\p\theta_{SD}}{\p\log\mu} = b\,. \ee
Hence, although the self-dual sector of QCD is perturbatively a CFT, indeed it has no coupling, non-perturbatively the self-dual $\theta$-angle flows according to the one-loop QCD $\beta$-function.

Our conventions are such that the path integral measure involves $e^{-S}$, where $S$ is the action above.  In the self-dual theory an instanton of charge $k$ is weighted by $e^{-\theta_{SD}k}$ in the path integral.  For an asymptotically free theory $b$ is positive, so that instantons are exponentially suppressed in the UV.

That the one-loop measure on instanton moduli space is determined by the one-loop $\beta$-function is well known \cite{tHooft:1976snw,Vainshtein:1981wh}.  In general it receives contributions from both from massive and zero-modes which conspire to reproduce \eqref{eqn:beta1}.  In supersymmetric theories the massive contributions cancel, and the one-loop coefficient can be obtained by counting zero-modes.  We will see that index-theoretic calculations on twistor space can recover the combined contribution in non-supersymmetric theories.


\section{Computing the Flow of the $\theta$-Angle Using the Index Theorem} \label{sec:index}

We will use the index theorem to give a different computation of the scale dependence of $\theta_{SD}$.  One-loop Feynman diagrams contribute \cite{tHooft:1976snw,Brown:1978yj,Vainshtein:1981wh} to the measure on the moduli space $\mc{M}^{inst}_k$ of instantons, and we will use the index theorem to compute how this determinant transforms under scaling of space-time. 

The failure of the determinant to be scale invariant is an anomaly.  Let us recall how anomalies for flavour symmetries can be computed using the index theorem, following \cite{Nielsen:1977aw}.  Consider some free fermions on $\R^4$ with a $G$ flavour symmetry.  We can study the theory in the presence of a background $G$-bundle by replacing the Dirac operator by the covariant Dirac operator.  If $P\to X$ is a principal $G$-bundle, then we get a family of theories parametrized by $X$.

The partition function of the theory is a section of a determinant line bundle on $X$. The curvature of the determinant line bundle can be computed using the family index theorem \cite{Atiyah:1970ws,Quillen:1985det,Bismut:1986ana}; this curvature represents the anomaly.

We will use this technique to study the anomaly to scale invariance.  Instead of considering a theory on $\R^4$ in the presence of a family of $G$-bundles parametrized by some $X$, we will consider instead some family of manifolds parametrized by $X$.  

We write $\PT = \Oo(1)^2 \to \CP^1$ for the twistor space of $\R^4$.  Let $X$ be some manifold equipped with a complex line bundle $L$, which we assume admits a square root.  Given this data, we can build a family of twistor spaces $\PT_L\to X$, as the total space of the vector bundle
\begin{equation}
    \Oo(1) \otimes L \oplus \Oo(1) \otimes L \to X \times \CP^1.
\end{equation}
The moduli space of charge $k$ instantons $\mc{M}^{inst}_k$ is the moduli of holomorphic bundles on $\PT$ with second Chern class $k$, framed at $\infty$ \cite{Ward:1977ta,atiyah1978construction}. Let $\mc{M}^{inst}_{k,L} \to X$ be the corresponding family over $X$ of moduli spaces of bundles on the fibres of $\PT_L \to X$.

The space of linearly embedded $\CP^1$s in $\PT$ is complexified space-time.  Correspondingly, we can build a family of complexified space-times from the family of twistor spaces $\PT_L$. This family of complexified space-times is simply $L \otimes \C^4$; it is clear that the scale is controlled by the line bundle $L$.

The determinant contribution to the measure on $\mc{M}^{inst}_{k,L}$ is a section of a determinant line bundle. We will find that the first Chern class of this line bundle, evaluated using the index theorem, is
\be bk\chern_1(L)\,. \ee
That is, the line bundle itself is isomorphic to $L^{\otimes bk}$. The anomaly to scale of the determinant of the $\dbar$ operator is $bk$, as desired.

Let us describe the index theory calculation we need to do. Writing
\be E \to \PT_L \times_X \mc{M}^{inst}_{k,L} \ee
for the universal bundle, positive helicity gauge bosons live in the Dolbeault complex of twistor space valued in the bundle $\mf{sl}(E)$.  Negative helicity gluons live in $\mf{sl}(E)\otimes K$, where $K = \Oo(-4)\otimes L^{-2}$ is the canonical bundle. Writing $\Pi$ for a parity shift, positive helicity Dirac fermions live in $\Pi(E\oplus E^\vee)\otimes\C^{N_f}\otimes K^{1/4}$ and negative helicity Dirac fermions live in $\Pi(E \oplus E^\vee) \otimes \C^{N_f} \otimes K^{3/4}$.  Fundamental complex scalars live in $(E \oplus E^\vee) \otimes K^{1/2} \otimes \C^{N_s}$.  The total field content lives in the bundle
\bea \label{eqn:R_instanton}
&R = \mf{sl}(E) \otimes (\Oo\oplus K) \\ 
&\oplus \Pi(E \oplus E^\vee) \otimes (K^{1/4} \oplus K^{3/4}) \otimes \C^{N_f} \\ 
&\oplus(E \oplus E^\vee) \otimes K^{1/2} \otimes \C^{N_s}\,.
\eea
We are interested in the square root of the determinant of the $\dbar$ operator valued in $R$.  We take the square root because the Dolbeault complex valued in $R$ describes fields in the BV formalism, including anti-fields, and the path integral measure in the BV setting is a section the square root of the canonical bundle of the space of fields \cite{Schwarz:1992nx}.

Let $T_{\PT_L / X}$ be the tangent bundle pointing along the fibres of the map $\PT_L \to X$. According to the Grothendieck-Riemann-Roch theorem \cite{borel1958theoreme,Hirzebruch:1966top,Grothendieck:1968cl} (a precursor of the family index theorem), the first Chern class of the determinant line is given by the integral
\be \frac{1}{2}\int_{\PT_L} \big[\Todd( T_{\PT_L/X} )\Cha(R) \big]_8\,. \ee
The factor of $\half$ appears because we work in the BV formalism. The integral is taken along the fibres of the map $\PT_L \times_X \mc{M}^{inst}_{k,L} \to \mc{M}^{inst}_{k,L}$, and $[ \ \ ]_8$ indicates we only take the cohomology class of degree $8$.  Note that the integrand here is the equivariant Todd class and the equivariant Chern class with equivariant parameter $q=\chern_1(L)$; our computation can be viewed as being an equivariant index theory calculation, instead of a family index theory calculation. (That the equivariant and family index theorems are related is well understood, see for example \cite{Freed:2016mpb}.)

Let us compute the Todd class of $T_{\PT_L/X}$.  There is a short exact sequence
\be 0 \to (\Oo(1) \otimes L )^2 \to T_{\PT_L/X} \to \Oo(2) \to 0\,. \ee
Therefore, the Todd class of $T_{\PT_L/X}$ is 
\be
\Todd(T_{\PT_L/X}) =  \op{Td}( \Oo(1) \otimes L)^2 \op{Td}(\Oo(2))\,.
\ee
Let $H$ be the first Chern class of $\Oo(1)$.  For any line bundle $S$, recall that the Todd class is 
\be \Todd(S) = 1 + \frac{\chern_1(S)}{2} + \frac{\chern_1(S)^2}{12} + \dots\,. \ee
Since $H^2 = 0$, and we are only interested in terms that are linear in $\chern_1(L)$,  only the first three terms of the Todd class will play a role.  We find that
\bea
&\Todd(T_{\PT_L/X}) \\
&= \Big(1 + \frac{1}{2} H + \frac{1}{2}\chern_1(L) + \frac{1}{6}\chern_1(L) H\Big)^2(1 + H) \\
&= \Big(1 + H + \chern_1(L) + \frac{5}{6}\chern_1(L) H\Big) (1+H) \\
&= 1 + 2 H + \chern_1(L) + \frac{11}{6}\chern_1(L) H\,.
\eea
Here we have dropped all terms involving $H^2$ or $\chern_1(L)^2$.   

Since $R$ is a real vector bundle, it has only even Chern characters.  Only the second and fourth Chern characters will play a role, as $\cha_0(R)$ will multiply $\op{td}_4(T_{\PT_L/X})$ which is zero. Because of this, the difference between $\mf{sl}(E)$ and $E^\vee \otimes E$ does not matter.  Further, terms in $\cha_4(R)$ can only contribute if they are divisible by $\chern_1(L)$.  To get a non-zero integral over $\PT_L$, we also need only retain those terms in $\Cha(R)$ which are divisible by $\cha_2(E)$.  Bearing this in mind, we can compute the contributions of each field to half of the Chern character of $R$.
\begin{itemize}
    \item The gauge boson bundle $\fsl(N_c)\otimes(\Oo\oplus K)$ contributes
    \be 2N_c\big(1-2H-\chern_1(L) + 4\chern_1(L)H\big)\cha_2(E)\,. \ee
    \item The Dirac fermion bundle $\Pi(E\oplus E^\vee)\otimes(K^{1/4}\oplus K^{3/4})\otimes\C^{N_f}$ contributes   \be  -2N_f\Big(1-2H-\chern_1(L) + \frac{5}{2}\chern_1(L)H\Big)\cha_2(E)\,. \ee
    \item The complex scalar bundle $(E\oplus E^\vee)\otimes K^{1/2}\otimes\C^{N_s}$ contributes
    \be N_s\big(1-2H-\chern_1(L) + 2\chern_1(L)H\big)\cha_2(E)\,. \ee
\end{itemize}
We need to multiply the sum of these three terms by the Todd class, retaining only those terms divisible by $\chern_1(L)H\cha_2(E)$.  The integral over $\PT_L$ will send $H\cha_2(E)$ to the instanton number $k$. 

The result is
\bea
&\bigg(\frac{11}{6}(2N_c - 2N_f + N_s) - 4(2N_c - 2N_f + N_s) \\
&+ (8N_c - 5N_f + 2N_s)\bigg)\chern_1(L)\int_{\PT_L}H\cha_2(E) \\
&= \bigg(\frac{11}{3}N_c - \frac{2}{3}N_f - \frac{1}{6}N_s\bigg)k\chern_1(L) 
\eea
matching precisely our expectation from equation \eqref{eqn:thetaSD_flow}.


\section{Weyl Anomaly on Gibbons-Hawking Spaces} \label{sec:GH}

Gravitational counterparts of the above techniques can also be used to compute one-loop Weyl anomalies.  Similarities between the twistorial anomalies of self-dual theories and one-loop Weyl anomalies have previously been noted in \cite{Doran:2023cmj}; here we will see that the latter arise as particular examples of the former \footnote{For example, the portion of the quantum effective action of a 4d field theory determined by the trace anomaly can be desribed by a dimension zero scalar field coupling to curvatures \cite{Riegert:1984kt}.  This is reminiscent of the Green-Schwarz mechanism for twistorial anomalies \cite{Costello:2021bah}.}.  As for the one-loop $\beta$-function, the Weyl anomalies of supersymmetric theories can be recovered by counting space-time zero-modes \cite{Christensen:1978md}.  In contrast, the index theorem on twistor space is sensitive to the contributions of massive modes in non-supersymmetric theories.

Consider self-dual QCD on the Einstein four-manifold $(M,g)$. Constant infinitesimal scale transformations act by $\delta g_{\mu\nu} = 2g_{\mu\nu}$, so the scale dependence of the partition function $\mc{Z}$ can be computed using
\be \int_M2g^{\mu\nu}\frac{\delta\log\mc{Z}}{\delta g^{\mu\nu}} = - \int_M\mrm{vol}_g\,g^{\mu\nu}\la T_{\mu\nu}\ra \ee
where $T_{\mu\nu} = - 2\delta S/\delta g^{\mu\nu}$ is the Hilbert stress-energy tensor. The right hand side is determined by the one-loop Weyl anomaly
\be \label{eq:Weyl-anomaly} g^{\mu\nu}\la T_{\mu\nu}\ra = \frac{1}{(4\pi)^2}\big(cC^2 - aE + \xi\triangle S\big) - \frac{b}{2(4\pi)^2}\mrm{tr}\big(F^2\big)\,, \ee
where $C^2 = C_{\mu\nu\rho\sigma}C^{\mu\nu\rho\sigma}$ for $C$ the Weyl tensor, $E$ is the Euler density and $S$ is the Ricci scalar \cite{Capper:1974ic,Duff:1977ay}. For an Einstein manifold $M$ these invariants simplify: $C^2 = E$ and $S=0$, so the constant $\xi$ drops out completely. Specialising to the case of self-dual QCD the anomaly coefficients are \cite{Duff:1993wm}
\bea \label{eqn:ac_coefficients}
&a = \frac{1}{180}\Big(31N_g + \frac{11}{2}N_fN_c + N_sN_c\Big)\,, \\
&c = \frac{1}{60}\big(6N_g + 3N_fN_c + N_sN_c\big)
\eea
for $N_g = \dim\gSU(N_c) = N_c^2-1$ the number of gluons. One-loop exactness of self-dual QCD ensures there are no higher loop corrections.

Now suppose we further demand that $(M,g)$ is a self-dual Einstein four-manifold, i.e., that both $\text{Ric}=0$ and $C = \ast C$. Then
\be \label{eqn:Pontryagin} \delta\log\mc{Z} = - \frac{1}{8\pi^2}(c-a)\int_M\op{tr}_{T_M}(C\wedge C) \ee
where we're interpreting $C$ as a 2-form with values in endomorphisms of the tangent bundle. We recognise the right hand side as $(c-a)p$ for $p$ the Pontryagin class of $M$.

We will use index theory to give an independent computation of \eqref{eqn:Pontryagin}. In order to do so we restrict to a simple class self-dual Einstein four-manifolds: the Gibbons-Hawking metrics \cite{Gibbons:1978tef}. These are asymptotically locally Euclidean, that is, they resemble $\R^4/\Gamma$ at infinity for some finite subgroup $\Gamma\subset\gSU(2)$. In the particular case of the Gibbons-Hawking metrics $\Gamma$ is the cyclic group of order $m\in\Z_{\geq2}$. Explicitly
\be \label{eqn:GH_metric} g_\text{GH} = \frac{(\dif\tau + \vec{\omega}(\vec{x})\cdot\dif\vec{x})^2}{V(\vec{x})} + V(\vec{x})\|\dif\vec{x}\|^2 \ee
where $\vec{x}\in\R^3$, $\tau\sim\tau+4\pi$ are co-ordinates, and $\vec{\nabla}\times\vec{\omega}(\vec{x}) = \vec V(\vec{x})$ for
\be V(\vec{x}) = \sum_{n=1}^m\frac{1}{\|\vec{x}-\vec{x}_n\|}\,. \ee
The metric depends on $m$ points $\{\vec{x}_n\}_{n=1}^m\in\R^{3m}$, though acting by the Euclidean group on these parameters yields isometric geometries. Therefore the moduli space of Gibbons-Hawking metrics $\cM^\text{GH}$ has dimension $3(m-2)$ (except in the case $m=2$ where it is one-dimensional due to axisymmetry). The metric \eqref{eqn:GH_metric} is non-singular as long as none of the centres $\vec{x}_n$ coincide. Constant scale transformations $s\in\R_{>0}$ acting by $g_\text{GH} \mapsto s^2g_\text{GH}$ induce the action $\vec{x}_n\mapsto s^2\vec{x}_n$ on $\cM^\text{GH}$.

The advantage of restricting to Gibbons-Hawking metrics is that their twistor spaces have a simple description due to Hitchin \cite{Hitchin:1979rts}.  Recall Penrose's non-linear graviton \cite{Penrose:1976js}: It identifies complex four-manifolds equipped with self-dual Ricci-flat holomorphic metrics and complex three-folds that fibre holomorphically over $\CP^1$ with an $\mscr{O}(2)$ valued symplectic form on the fibres. (The complex three-fold is also required to admit a rational curve with normal bundle $\mscr{O}(1)\oplus\mscr{O}(1)$.)  To recover a self-dual Ricci-flat Riemannian four-manifold the twistor space must also be equipped with an anti-holomorphic involution $\sigma$ acting as the antipodal map on the rational curve.

The twistor spaces of the Gibbons-Hawking metrics \eqref{eqn:GH_metric} are the hypersurfaces in $\mscr{O}(m)\oplus\mscr{O}(m)\oplus\mscr{O}(2)\to\CP^1$ cut out by
\be \label{eqn:GH_twistor} XY = \prod_{n=1}^m\big(Z^2 - \vec{x}_n(\zeta)\big)\,, \ee
where the points in $\vec{x}_n\in\R^3$ have been identified with sections $\vec{x}_n(\zeta)$ of $\mscr{O}(2)\to\CP^1$ equivariant under $(Z,\zeta)\mapsto(\bar Z/\bar\zeta^2,\zeta\mapsto -1/\bar\zeta)$. The twistor spaces \eqref{eqn:GH_twistor} admit an anti-holomorphic involution
\be \sigma:(X,Y,Z,\zeta)\mapsto(\bar X/\bar\zeta^m,\bar Y/\bar\zeta^m,\bar Z/\bar\zeta^2,-1/\bar\zeta) \ee
encoding the Riemannian slice.  Forgetting the involution $\sigma$, the moduli space $\cM^\text{GH}$ naturally complexifies to $\cM^\text{GH}_\C$ where the points $\{\vec{x}_n\}_{n=1}^m$ are now permitted to take values in $\C^{3m}$. Sitting above this moduli space is a universal bundle of twistor spaces $\cPT\to\cM^\text{GH}_\C$.  Scale transformations complexify to a $\C^\times$ action
\be \label{eqn:complexified_action} (X,Y,Z,\zeta,\vec{x}_n)\mapsto (sX,sY,sZ,\zeta,s^2\vec{x}_n) \ee
on the universal bundle.  We will now compute the anomaly to this $\C^\times$ action.


\section{Computing the Weyl Anomaly using the Index Theorem} \label{sec:Weyl}

Our strategy is to mimic the calculation in Section \ref{sec:index}. We can define self-dual QCD at points of $\cM^\text{GH}_\C$ as a holomorphic theory on the associated twistor space. We will study a family of self-dual gauge theories on complexified Gibbons-Hawking spaces whose scale is controlled by the line bundle $L\to X$. We begin by forming the associated bundle $\cPT_L = L\times_{\C^\times}\cPT\to X$. This factors through the family of moduli spaces $\cM_L^\text{GH} = L\times_{\C^\times}\cM^\text{GH}_\C\to X$. As in the case of instantons, we consider the square root of the determinant of the $\dbar$ operator on a modification of the bundle $R$ \eqref{eqn:R_instanton}: we replace the bundle $E$ with the trivial $\C^{N_c}$ bundle over $\cPT_L$, $\Oo$ by the structure sheaf of $\cPT_L$ and take $K = \mrm{Det}(T_{\cPT_L/\cM^\text{GH}_L})^{-1}$.

The partition function of holomorphic BF theory on this family is a section of the determinant line bundle over $\cM_L^\text{GH}$. Applying the Grothendieck-Riemann-Roch theorem the first Chern class of the determinant line is
\be \label{eqn:GH_index} \frac{1}{2}\int_{\cPT_L}\Big[\Todd\big(T_{\cPT_L/\cM^\text{GH}_L}\big)\Cha(R))\Big]_8\,. \ee
The integral is taken along the fibres of the map $\cPT_L\to\cM_L^\text{GH}$.

In order to evaluate \eqref{eqn:GH_index} we exploit the fact that the twistor space of a self-dual Einstein manifold fibres holomorphically over $\CP^1$. We therefore have a holomorphic projection map $\pi:\cPT\to\CP^1\times\cM^\text{GH}_\C$, and differentiating yields the short exact sequence
\be 0\to\cN\to T_{\cPT/\cM^\text{GH}_\C}\to T_{\CP^1}\to 0 \ee
where we abuse notation by writing $T_{\CP^1}$ for its pullback to $\cPT$ and $\cN$ denotes the normal bundle to the $\CP^1$ in the twistor space directions. From the complexified scaling action \eqref{eqn:complexified_action} this short exact sequence is $\C^\times$ equivariant if we assign $\cN$ charge one. Forming the associated bundles gives the following short exact sequence over $\cPT_L$
\be \label{eqn:additivity} 0\to L\otimes\cN\to T_{\cPT_L/\cM^\text{GH}_L}\to T_{\CP^1}\to 0\,, \ee
allowing the index \eqref{eqn:GH_index} to be evaluated by additivity; the computation is included in the Supplementary Material \ref{supp:GH_index}. \nocite{supplemental} We find the characteristic classes evaluate to
\begin{widetext}
\bea \label{eqn:evaluate_index}
&- \frac{1}{12}\bigg(\frac{26}{15}\big(\op{rk}(\mf{sl}(E)) - 2N_f\op{rk}(E) + N_s\op{rk}(E)\big) 
 - 4\big(\op{rk}(\mf{sl}(E)) - 2N_f\op{rk}(E) + N_s\op{rk}(E)\big) 
\\ 
&+  \big(4\op{rk}(\mf{sl}(E)) 
- 5N_f\op{rk}(E) + 2N_s\op{rk}(E)\big)\bigg)\chern_1(L)\int_{\cPT_L}H\cha_2(\cN) \\
&= - \frac{1}{180}\big(26N_g - 7N_fN_c - 
4N_sN_c\big)\chern_1(L)\int_{\cPT_L}H\cha_2(\cN)\,.
\eea
\end{widetext}
Choosing $H$ to have support at a point $\zeta\in\CP^1$ the integral localises to the fibre $\pi^{-1}(\zeta) = M_\text{GH}(\zeta)$: the Gibbons-Hawking space in the complex structure corresponding to $\zeta$. The normal bundle $\cN$ can be identified with $T^{1,0}M_\text{GH}(\zeta)$, and its second Chern character is independent of $\zeta$.
\bea
&\cha_2(T^{1,0}M_\text{GH}(\zeta)) = - \frac{1}{2}\chern_2(T^{1,0}M_\text{GH}(\zeta)\oplus T^{0,1}M_\text{GH}(\zeta)) \\
&= - \frac{1}{2}\chern_2(T_\C M_\text{GH}) = \frac{1}{2}p\,. 
\eea
Here $M_\text{GH}$ is the Gibbons-Hawking space at a particular point of the moduli space $\cM^\text{GH}_L$. We therefore find that the index evaluates to
\bea
- \frac{1}{360}\big(26N_g - 7N_fN_c - 4N_sN_c\big)\chern_1(L)p\,. 
\eea
Comparing to the $a,c$ coefficients in equation \eqref{eqn:ac_coefficients} we recognise this as $(c-a)p$. Equivalently, the determinant line bundle is isomorphic $L^{\otimes(c-a)p}$ matching our expectation from equation \eqref{eqn:Pontryagin}.


\section{Discussion}

We've evaluated the one-loop QCD $\beta$-function using index-theoretic arguments on twistor space.  Can higher-loop contributions to the $\beta$-function can also be obtained? In \cite{Costello:2021bah} one of the authors demonstrated a two-loop flow in the operator $\op{tr}(B^2)$ of the form
\be \frac{\p}{\p\log\mu}\op{tr}(B\wedge B) = - \frac{\lambda^2}{24\pi^4}\op{tr}(F\wedge F) \ee
under the assumption that
\be \op{tr}_{\mrm{Ad}\oplus\Pi(\mrm{F}\oplus\mrm{F}^\vee)^{\oplus N_f}\oplus\mrm{F}^{\oplus N_s}}(X^4) = \lambda^2\op{tr}_\mrm{F}(X^2)^2\,. \ee
Deforming self-dual QCD with this operator generates a two-loop contribution to the flow of the self-dual $\theta$-angle
\be \frac{\p}{\p\log\mu}\theta_\text{SD} = b - \frac{\lambda^2}{3}\frac{g^2_\text{YM}}{8\pi^2} \ee
which by the arguments of Section \ref{sec:theta_angle} is compensated by a two-loop flow of the QCD coupling:
\be \label{eqn:two_loop} \frac{\p}{\p\log\mu}g_\text{YM} = - b\frac{g^3_\text{YM}}{16\pi^2} + \frac{2\lambda^2}{3}\frac{g^5_\text{YM}}{(16\pi^2)^2} \ee
Beyond two loops we expect higher-order corrections to the above generated by multiple insertions of $\op{tr}(B^2)$.

Equation \eqref{eqn:two_loop} does not coincide with the familiar $\beta$-function at two loops \cite{Caswell:1974gg,Jones:1974mm,Jones:1974pg} \footnote{The two-loop flow of the canonical coupling also receives contributions from Yukawa couplings which are difficult to incorporate on twistor space \cite{Jones:1981we,Machacek:1983tz}.}. Indeed, $\lambda^2 = 0$ for $\cN=1$ supersymmetric Yang-Mills theory but the $\beta$-function has a non-vanishing two-loop contribution.  This discrepancy is accounted for by noting that $8\pi^2/g^2_\text{YM}$ is the real part of the `holomorphic' coupling $\theta_\text{SD}$, which is one-loop exact in supersymmetric gauge theory \cite{Novikov:1985rd,Arkani-Hamed:1997qui}. Rescaling the gauge field in order to canonically normalize the gauge kinetic term generates an anomalous Jacobian responsible for higher-loop contributions to the flow of the canonical coupling.

We have seen that the linear combination of anomaly coefficients $a-c$ can be computed via index theory. Can the two coefficients be recovered independently?  This should follow from applying the arguments of Section \ref{sec:Weyl} on a self-dual Riemannian four-manifold with non-vanishing Ricci tensor. A natural asymptotically flat candidate is Burns space \cite{lebrun1991explicit}.

Indeed, that twistorial anomalies are sensitive to the $a,c$ coefficients independently can be seen indirectly.  Infinitesimal conformal symmetries of a self-dual conformal background lift to holomorphic vector fields on the corresponding twistor space.  One-loop conformal anomalies therefore appear as (a subset of the) gravitational anomalies on twistor space.  These are polynomials in the Chern classes of the tangent bundle of degree eight, i.e., a linear combination of $\chern_1^4$, $\chern_2\chern_1^2$, $\chern_3\chern_1$ and $\chern_2^2$.  By evaluating the contribution from ordinary fields (conformally coupled scalars, fermions, gluons) and also from higher derivative fields appearing in the $\cN=4$ superconformal gravity supermultiplet (conformal gravitons, conformal gravitinos, antisymmetric tensors, higher derivative Weyl fermions and conformally coupled dimension zero complex scalars) we find that the four classes naturally organise into
\bea \label{eqn:sdcg_anomaly}
&\frac{1}{128}(16\chern_2-5\chern_1^2)\chern_1^2\,a + \frac{1}{64}(3\chern_1^2-8\chern_2)\chern_1^2\,c \\
&+ \frac{1}{46080}(96\chern_2^2-80\chern_1^2\chern_2+32\chern_1\chern_3+15\chern_1^4)\,\chi \\
&+ \frac{1}{64}(3\chern_1^3 - 16\chern_1\chern_2 + 64\chern_3)\chern_1\,\psi\,. \eea
Here $a,c$ are the Weyl anomaly coefficients.  The classes with coefficients $\chi,\psi$ do not represent space-time anomalies: $\chi$ is the coefficient of the twistorial anomaly in self-dual Einstein gravity which counts on-shell degrees of freedom weighted by parity \cite{Bittleston:2022nfr} and $\psi$ represents a novel twistorial anomaly in self-dual conformal gravity \footnote{Although we note that the twistorial anomaly coefficient $\chi$ is proportional to the one-loop vacuum energy \cite{Boyle:2021jaz}. It's not clear to us if $\psi$ also has a space-time avatar.}.  Values of the coefficients $a,c,\chi$ and $\psi$ are included in the Supplementary Material \ref{supp:sdcg-anomaly}.  In total we've evaluated the anomaly polynomial on eight different types of field; four are needed to fix the classes multiplying $a,c$, and the remaining four are non-trivial checks on the validity of \eqref{eqn:sdcg_anomaly}.


\begin{acknowledgments}
RB would like to thank Latham Boyle, Simon Heuveline, David Skinner, Neil Turok and Vatsalya Vaibhav for helpful conversations. The authors gratefully acknowledge the support of the Simons Collaboration on Celestial Holography. Research at Perimeter Institute is supported by the Government of Canada through Industry Canada and by the Province of Ontario through the Ministry of Research and Innovation.
\end{acknowledgments}


\bibliography{main}

\begin{thebibliography}{51}%
\makeatletter
\providecommand \@ifxundefined [1]{%
 \@ifx{#1\undefined}
}%
\providecommand \@ifnum [1]{%
 \ifnum #1\expandafter \@firstoftwo
 \else \expandafter \@secondoftwo
 \fi
}%
\providecommand \@ifx [1]{%
 \ifx #1\expandafter \@firstoftwo
 \else \expandafter \@secondoftwo
 \fi
}%
\providecommand \natexlab [1]{#1}%
\providecommand \enquote  [1]{``#1''}%
\providecommand \bibnamefont  [1]{#1}%
\providecommand \bibfnamefont [1]{#1}%
\providecommand \citenamefont [1]{#1}%
\providecommand \href@noop [0]{\@secondoftwo}%
\providecommand \href [0]{\begingroup \@sanitize@url \@href}%
\providecommand \@href[1]{\@@startlink{#1}\@@href}%
\providecommand \@@href[1]{\endgroup#1\@@endlink}%
\providecommand \@sanitize@url [0]{\catcode `\\12\catcode `\$12\catcode `\&12\catcode `\#12\catcode `\^12\catcode `\_12\catcode `\%12\relax}%
\providecommand \@@startlink[1]{}%
\providecommand \@@endlink[0]{}%
\providecommand \url  [0]{\begingroup\@sanitize@url \@url }%
\providecommand \@url [1]{\endgroup\@href {#1}{\urlprefix }}%
\providecommand \urlprefix  [0]{URL }%
\providecommand \Eprint [0]{\href }%
\providecommand \doibase [0]{http://dx.doi.org/}%
\providecommand \selectlanguage [0]{\@gobble}%
\providecommand \bibinfo  [0]{\@secondoftwo}%
\providecommand \bibfield  [0]{\@secondoftwo}%
\providecommand \translation [1]{[#1]}%
\providecommand \BibitemOpen [0]{}%
\providecommand \bibitemStop [0]{}%
\providecommand \bibitemNoStop [0]{.\EOS\space}%
\providecommand \EOS [0]{\spacefactor3000\relax}%
\providecommand \BibitemShut  [1]{\csname bibitem#1\endcsname}%
\let\auto@bib@innerbib\@empty
\bibitem [{\citenamefont {Politzer}(1973)}]{Politzer:1973fx}%
  \BibitemOpen
  \bibfield  {author} {\bibinfo {author} {\bibfnamefont {H.~D.}\ \bibnamefont {Politzer}},\ }\href {\doibase 10.1103/PhysRevLett.30.1346} {\bibfield  {journal} {\bibinfo  {journal} {Phys. Rev. Lett.}\ }\textbf {\bibinfo {volume} {30}},\ \bibinfo {pages} {1346} (\bibinfo {year} {1973})}\BibitemShut {NoStop}%
\bibitem [{\citenamefont {Gross}\ and\ \citenamefont {Wilczek}(1973)}]{Gross:1973ju}%
  \BibitemOpen
  \bibfield  {author} {\bibinfo {author} {\bibfnamefont {D.~J.}\ \bibnamefont {Gross}}\ and\ \bibinfo {author} {\bibfnamefont {F.}~\bibnamefont {Wilczek}},\ }\href {\doibase 10.1103/PhysRevD.8.3633} {\bibfield  {journal} {\bibinfo  {journal} {Phys. Rev. D}\ }\textbf {\bibinfo {volume} {8}},\ \bibinfo {pages} {3633} (\bibinfo {year} {1973})}\BibitemShut {NoStop}%
\bibitem [{\citenamefont {Abbott}(1981)}]{Abbott:1980hw}%
  \BibitemOpen
  \bibfield  {author} {\bibinfo {author} {\bibfnamefont {L.~F.}\ \bibnamefont {Abbott}},\ }\href {\doibase 10.1016/0550-3213(81)90371-0} {\bibfield  {journal} {\bibinfo  {journal} {Nucl. Phys. B}\ }\textbf {\bibinfo {volume} {185}},\ \bibinfo {pages} {189} (\bibinfo {year} {1981})}\BibitemShut {NoStop}%
\bibitem [{\citenamefont {Abbott}(1982)}]{Abbott:1981ke}%
  \BibitemOpen
  \bibfield  {author} {\bibinfo {author} {\bibfnamefont {L.~F.}\ \bibnamefont {Abbott}},\ }\href@noop {} {\bibfield  {journal} {\bibinfo  {journal} {Acta Phys. Polon. B}\ }\textbf {\bibinfo {volume} {13}},\ \bibinfo {pages} {33} (\bibinfo {year} {1982})}\BibitemShut {NoStop}%
\bibitem [{\citenamefont {Chalmers}\ and\ \citenamefont {Siegel}(1996)}]{Chalmers:1996rq}%
  \BibitemOpen
  \bibfield  {author} {\bibinfo {author} {\bibfnamefont {G.}~\bibnamefont {Chalmers}}\ and\ \bibinfo {author} {\bibfnamefont {W.}~\bibnamefont {Siegel}},\ }\href {\doibase 10.1103/PhysRevD.54.7628} {\bibfield  {journal} {\bibinfo  {journal} {Phys. Rev. D}\ }\textbf {\bibinfo {volume} {54}},\ \bibinfo {pages} {7628} (\bibinfo {year} {1996})},\ \Eprint {http://arxiv.org/abs/hep-th/9606061} {arXiv:hep-th/9606061} \BibitemShut {NoStop}%
\bibitem [{\citenamefont {Losev}\ \emph {et~al.}(2018)\citenamefont {Losev}, \citenamefont {Polyubin},\ and\ \citenamefont {Rosly}}]{Losev:2017qrj}%
  \BibitemOpen
  \bibfield  {author} {\bibinfo {author} {\bibfnamefont {A.}~\bibnamefont {Losev}}, \bibinfo {author} {\bibfnamefont {I.}~\bibnamefont {Polyubin}}, \ and\ \bibinfo {author} {\bibfnamefont {A.}~\bibnamefont {Rosly}},\ }\href {\doibase 10.1007/JHEP02(2018)041} {\bibfield  {journal} {\bibinfo  {journal} {JHEP}\ }\textbf {\bibinfo {volume} {02}},\ \bibinfo {pages} {041} (\bibinfo {year} {2018})},\ \Eprint {http://arxiv.org/abs/1711.10026} {arXiv:1711.10026 [hep-th]} \BibitemShut {NoStop}%
\bibitem [{\citenamefont {Sch\"afer}\ and\ \citenamefont {Shuryak}(1998)}]{Schafer:1996wv}%
  \BibitemOpen
  \bibfield  {author} {\bibinfo {author} {\bibfnamefont {T.}~\bibnamefont {Sch\"afer}}\ and\ \bibinfo {author} {\bibfnamefont {E.~V.}\ \bibnamefont {Shuryak}},\ }\href {\doibase 10.1103/RevModPhys.70.323} {\bibfield  {journal} {\bibinfo  {journal} {Rev. Mod. Phys.}\ }\textbf {\bibinfo {volume} {70}},\ \bibinfo {pages} {323} (\bibinfo {year} {1998})},\ \Eprint {http://arxiv.org/abs/hep-ph/9610451} {arXiv:hep-ph/9610451} \BibitemShut {NoStop}%
\bibitem [{Note1()}]{Note1}%
  \BibitemOpen
  \bibinfo {note} {The flow of $\theta _{SD}$ was computed by a direct Feynman diagram analysis in \cite {Losev:2017qrj}.}\BibitemShut {Stop}%
\bibitem [{\citenamefont {'t~Hooft}(1976)}]{tHooft:1976snw}%
  \BibitemOpen
  \bibfield  {author} {\bibinfo {author} {\bibfnamefont {G.}~\bibnamefont {'t~Hooft}},\ }\href {\doibase 10.1103/PhysRevD.14.3432} {\bibfield  {journal} {\bibinfo  {journal} {Phys. Rev. D}\ }\textbf {\bibinfo {volume} {14}},\ \bibinfo {pages} {3432} (\bibinfo {year} {1976})},\ \bibinfo {note} {[Erratum: Phys.Rev.D 18, 2199 (1978)]}\BibitemShut {NoStop}%
\bibitem [{\citenamefont {Vainshtein}\ \emph {et~al.}(1982)\citenamefont {Vainshtein}, \citenamefont {Zakharov}, \citenamefont {Novikov},\ and\ \citenamefont {Shifman}}]{Vainshtein:1981wh}%
  \BibitemOpen
  \bibfield  {author} {\bibinfo {author} {\bibfnamefont {A.~I.}\ \bibnamefont {Vainshtein}}, \bibinfo {author} {\bibfnamefont {V.~I.}\ \bibnamefont {Zakharov}}, \bibinfo {author} {\bibfnamefont {V.~A.}\ \bibnamefont {Novikov}}, \ and\ \bibinfo {author} {\bibfnamefont {M.~A.}\ \bibnamefont {Shifman}},\ }\href {\doibase 10.1070/PU1982v025n04ABEH004533} {\bibfield  {journal} {\bibinfo  {journal} {Sov. Phys. Usp.}\ }\textbf {\bibinfo {volume} {25}},\ \bibinfo {pages} {195} (\bibinfo {year} {1982})}\BibitemShut {NoStop}%
\bibitem [{\citenamefont {Brown}\ and\ \citenamefont {Creamer}(1978)}]{Brown:1978yj}%
  \BibitemOpen
  \bibfield  {author} {\bibinfo {author} {\bibfnamefont {L.~S.}\ \bibnamefont {Brown}}\ and\ \bibinfo {author} {\bibfnamefont {D.~B.}\ \bibnamefont {Creamer}},\ }\href {\doibase 10.1103/PhysRevD.18.3695} {\bibfield  {journal} {\bibinfo  {journal} {Phys. Rev. D}\ }\textbf {\bibinfo {volume} {18}},\ \bibinfo {pages} {3695} (\bibinfo {year} {1978})}\BibitemShut {NoStop}%
\bibitem [{\citenamefont {Nielsen}\ and\ \citenamefont {Schroer}(1977)}]{Nielsen:1977aw}%
  \BibitemOpen
  \bibfield  {author} {\bibinfo {author} {\bibfnamefont {N.~K.}\ \bibnamefont {Nielsen}}\ and\ \bibinfo {author} {\bibfnamefont {B.}~\bibnamefont {Schroer}},\ }\href {\doibase 10.1016/0550-3213(77)90453-9} {\bibfield  {journal} {\bibinfo  {journal} {Nucl. Phys. B}\ }\textbf {\bibinfo {volume} {127}},\ \bibinfo {pages} {493} (\bibinfo {year} {1977})}\BibitemShut {NoStop}%
\bibitem [{\citenamefont {Atiyah}\ and\ \citenamefont {Singer}(1971)}]{Atiyah:1970ws}%
  \BibitemOpen
  \bibfield  {author} {\bibinfo {author} {\bibfnamefont {M.~F.}\ \bibnamefont {Atiyah}}\ and\ \bibinfo {author} {\bibfnamefont {I.~M.}\ \bibnamefont {Singer}},\ }\href {\doibase 10.2307/1970756} {\bibfield  {journal} {\bibinfo  {journal} {Annals Math.}\ }\textbf {\bibinfo {volume} {93}},\ \bibinfo {pages} {119} (\bibinfo {year} {1971})}\BibitemShut {NoStop}%
\bibitem [{\citenamefont {Quillen}(1985)}]{Quillen:1985det}%
  \BibitemOpen
  \bibfield  {author} {\bibinfo {author} {\bibfnamefont {D.}~\bibnamefont {Quillen}},\ }\href@noop {} {\bibfield  {journal} {\bibinfo  {journal} {Functional Analysis and its Applications}\ }\textbf {\bibinfo {volume} {19}},\ \bibinfo {pages} {31} (\bibinfo {year} {1985})}\BibitemShut {NoStop}%
\bibitem [{\citenamefont {Bismut}\ and\ \citenamefont {Freed}(1986)}]{Bismut:1986ana}%
  \BibitemOpen
  \bibfield  {author} {\bibinfo {author} {\bibfnamefont {J.-M.}\ \bibnamefont {Bismut}}\ and\ \bibinfo {author} {\bibfnamefont {D.~S.}\ \bibnamefont {Freed}},\ }\href@noop {} {\bibfield  {journal} {\bibinfo  {journal} {Communications in mathematical physics}\ }\textbf {\bibinfo {volume} {106}},\ \bibinfo {pages} {159} (\bibinfo {year} {1986})}\BibitemShut {NoStop}%
\bibitem [{\citenamefont {Ward}(1977)}]{Ward:1977ta}%
  \BibitemOpen
  \bibfield  {author} {\bibinfo {author} {\bibfnamefont {R.~S.}\ \bibnamefont {Ward}},\ }\href {\doibase 10.1016/0375-9601(77)90842-8} {\bibfield  {journal} {\bibinfo  {journal} {Phys. Lett. A}\ }\textbf {\bibinfo {volume} {61}},\ \bibinfo {pages} {81} (\bibinfo {year} {1977})}\BibitemShut {NoStop}%
\bibitem [{\citenamefont {Atiyah}\ \emph {et~al.}(1978)\citenamefont {Atiyah}, \citenamefont {Hitchin}, \citenamefont {Drinfeld},\ and\ \citenamefont {Manin}}]{atiyah1978construction}%
  \BibitemOpen
  \bibfield  {author} {\bibinfo {author} {\bibfnamefont {M.}~\bibnamefont {Atiyah}}, \bibinfo {author} {\bibfnamefont {N.}~\bibnamefont {Hitchin}}, \bibinfo {author} {\bibfnamefont {V.}~\bibnamefont {Drinfeld}}, \ and\ \bibinfo {author} {\bibfnamefont {Y.~I.}\ \bibnamefont {Manin}},\ }\href@noop {} {\bibfield  {journal} {\bibinfo  {journal} {Physics Letters. A}\ }\textbf {\bibinfo {volume} {65}},\ \bibinfo {pages} {185} (\bibinfo {year} {1978})}\BibitemShut {NoStop}%
\bibitem [{\citenamefont {Schwarz}(1993)}]{Schwarz:1992nx}%
  \BibitemOpen
  \bibfield  {author} {\bibinfo {author} {\bibfnamefont {A.~S.}\ \bibnamefont {Schwarz}},\ }\href {\doibase 10.1007/BF02097392} {\bibfield  {journal} {\bibinfo  {journal} {Commun. Math. Phys.}\ }\textbf {\bibinfo {volume} {155}},\ \bibinfo {pages} {249} (\bibinfo {year} {1993})},\ \Eprint {http://arxiv.org/abs/hep-th/9205088} {arXiv:hep-th/9205088} \BibitemShut {NoStop}%
\bibitem [{\citenamefont {Borel}\ and\ \citenamefont {Serre}(1958)}]{borel1958theoreme}%
  \BibitemOpen
  \bibfield  {author} {\bibinfo {author} {\bibfnamefont {A.}~\bibnamefont {Borel}}\ and\ \bibinfo {author} {\bibfnamefont {J.-P.}\ \bibnamefont {Serre}},\ }\href@noop {} {\bibfield  {journal} {\bibinfo  {journal} {Bulletin de la Soci{\'e}t{\'e} math{\'e}matique de France}\ }\textbf {\bibinfo {volume} {86}},\ \bibinfo {pages} {97} (\bibinfo {year} {1958})}\BibitemShut {NoStop}%
\bibitem [{\citenamefont {Hirzebruch}\ \emph {et~al.}(1966)\citenamefont {Hirzebruch}, \citenamefont {Borel},\ and\ \citenamefont {Schwarzenberger}}]{Hirzebruch:1966top}%
  \BibitemOpen
  \bibfield  {author} {\bibinfo {author} {\bibfnamefont {F.}~\bibnamefont {Hirzebruch}}, \bibinfo {author} {\bibfnamefont {A.}~\bibnamefont {Borel}}, \ and\ \bibinfo {author} {\bibfnamefont {R.}~\bibnamefont {Schwarzenberger}},\ }\href@noop {} {\emph {\bibinfo {title} {Topological Methods in Algebraic Geometry}}},\ Vol.\ \bibinfo {volume} {175}\ (\bibinfo  {publisher} {Springer Berlin-Heidelberg-New York},\ \bibinfo {year} {1966})\BibitemShut {NoStop}%
\bibitem [{\citenamefont {Grothendieck}(1968)}]{Grothendieck:1968cl}%
  \BibitemOpen
  \bibfield  {author} {\bibinfo {author} {\bibfnamefont {A.}~\bibnamefont {Grothendieck}},\ }\href@noop {} {\bibfield  {journal} {\bibinfo  {journal} {SGA 6}\ } (\bibinfo {year} {1968})}\BibitemShut {NoStop}%
\bibitem [{\citenamefont {Freed}(2017)}]{Freed:2016mpb}%
  \BibitemOpen
  \bibfield  {author} {\bibinfo {author} {\bibfnamefont {D.~S.}\ \bibnamefont {Freed}},\ }\href {\doibase 10.4310/sdg.2017.v22.n1.a5} {\bibfield  {journal} {\bibinfo  {journal} {Surveys Diff. Geom.}\ }\textbf {\bibinfo {volume} {22}},\ \bibinfo {pages} {125} (\bibinfo {year} {2017})},\ \Eprint {http://arxiv.org/abs/1606.01129} {arXiv:1606.01129 [math.DG]} \BibitemShut {NoStop}%
\bibitem [{\citenamefont {Doran}\ \emph {et~al.}(2024)\citenamefont {Doran}, \citenamefont {Monteiro},\ and\ \citenamefont {Wikeley}}]{Doran:2023cmj}%
  \BibitemOpen
  \bibfield  {author} {\bibinfo {author} {\bibfnamefont {G.}~\bibnamefont {Doran}}, \bibinfo {author} {\bibfnamefont {R.}~\bibnamefont {Monteiro}}, \ and\ \bibinfo {author} {\bibfnamefont {S.}~\bibnamefont {Wikeley}},\ }\href {\doibase 10.1007/JHEP07(2024)139} {\bibfield  {journal} {\bibinfo  {journal} {JHEP}\ }\textbf {\bibinfo {volume} {07}},\ \bibinfo {pages} {139} (\bibinfo {year} {2024})},\ \Eprint {http://arxiv.org/abs/2312.13267} {arXiv:2312.13267 [hep-th]} \BibitemShut {NoStop}%
\bibitem [{Note2()}]{Note2}%
  \BibitemOpen
  \bibinfo {note} {For example, the portion of the quantum effective action of a 4d field theory determined by the trace anomaly can be desribed by a dimension zero scalar field coupling to curvatures \cite {Riegert:1984kt}. This is reminiscent of the Green-Schwarz mechanism for twistorial anomalies \cite {Costello:2021bah}.}\BibitemShut {Stop}%
\bibitem [{\citenamefont {Christensen}\ and\ \citenamefont {Duff}(1979)}]{Christensen:1978md}%
  \BibitemOpen
  \bibfield  {author} {\bibinfo {author} {\bibfnamefont {S.~M.}\ \bibnamefont {Christensen}}\ and\ \bibinfo {author} {\bibfnamefont {M.~J.}\ \bibnamefont {Duff}},\ }\href {\doibase 10.1016/0550-3213(79)90516-9} {\bibfield  {journal} {\bibinfo  {journal} {Nucl. Phys. B}\ }\textbf {\bibinfo {volume} {154}},\ \bibinfo {pages} {301} (\bibinfo {year} {1979})}\BibitemShut {NoStop}%
\bibitem [{\citenamefont {Capper}\ and\ \citenamefont {Duff}(1974)}]{Capper:1974ic}%
  \BibitemOpen
  \bibfield  {author} {\bibinfo {author} {\bibfnamefont {D.~M.}\ \bibnamefont {Capper}}\ and\ \bibinfo {author} {\bibfnamefont {M.~J.}\ \bibnamefont {Duff}},\ }\href {\doibase 10.1007/BF02748300} {\bibfield  {journal} {\bibinfo  {journal} {Nuovo Cim. A}\ }\textbf {\bibinfo {volume} {23}},\ \bibinfo {pages} {173} (\bibinfo {year} {1974})}\BibitemShut {NoStop}%
\bibitem [{\citenamefont {Duff}(1977)}]{Duff:1977ay}%
  \BibitemOpen
  \bibfield  {author} {\bibinfo {author} {\bibfnamefont {M.~J.}\ \bibnamefont {Duff}},\ }\href {\doibase 10.1016/0550-3213(77)90410-2} {\bibfield  {journal} {\bibinfo  {journal} {Nucl. Phys. B}\ }\textbf {\bibinfo {volume} {125}},\ \bibinfo {pages} {334} (\bibinfo {year} {1977})}\BibitemShut {NoStop}%
\bibitem [{\citenamefont {Duff}(1994)}]{Duff:1993wm}%
  \BibitemOpen
  \bibfield  {author} {\bibinfo {author} {\bibfnamefont {M.~J.}\ \bibnamefont {Duff}},\ }\href {\doibase 10.1088/0264-9381/11/6/004} {\bibfield  {journal} {\bibinfo  {journal} {Class. Quant. Grav.}\ }\textbf {\bibinfo {volume} {11}},\ \bibinfo {pages} {1387} (\bibinfo {year} {1994})},\ \Eprint {http://arxiv.org/abs/hep-th/9308075} {arXiv:hep-th/9308075} \BibitemShut {NoStop}%
\bibitem [{\citenamefont {Gibbons}\ and\ \citenamefont {Hawking}(1978)}]{Gibbons:1978tef}%
  \BibitemOpen
  \bibfield  {author} {\bibinfo {author} {\bibfnamefont {G.~W.}\ \bibnamefont {Gibbons}}\ and\ \bibinfo {author} {\bibfnamefont {S.~W.}\ \bibnamefont {Hawking}},\ }\href {\doibase 10.1016/0370-2693(78)90478-1} {\bibfield  {journal} {\bibinfo  {journal} {Phys. Lett. B}\ }\textbf {\bibinfo {volume} {78}},\ \bibinfo {pages} {430} (\bibinfo {year} {1978})}\BibitemShut {NoStop}%
\bibitem [{\citenamefont {Hitchin}(1979)}]{Hitchin:1979rts}%
  \BibitemOpen
  \bibfield  {author} {\bibinfo {author} {\bibfnamefont {N.~J.}\ \bibnamefont {Hitchin}},\ }\href {\doibase 10.1017/S0305004100055924} {\bibfield  {journal} {\bibinfo  {journal} {Math. Proc. Cambridge Phil. Soc.}\ }\textbf {\bibinfo {volume} {85}},\ \bibinfo {pages} {465} (\bibinfo {year} {1979})}\BibitemShut {NoStop}%
\bibitem [{\citenamefont {Penrose}(1976)}]{Penrose:1976js}%
  \BibitemOpen
  \bibfield  {author} {\bibinfo {author} {\bibfnamefont {R.}~\bibnamefont {Penrose}},\ }\href {\doibase 10.1007/BF00762011} {\bibfield  {journal} {\bibinfo  {journal} {Gen. Rel. Grav.}\ }\textbf {\bibinfo {volume} {7}},\ \bibinfo {pages} {31} (\bibinfo {year} {1976})}\BibitemShut {NoStop}%
\bibitem [{sup()}]{supplemental}%
  \BibitemOpen
  \href@noop {} {}\bibinfo {note} {See Supplemental Material at [URL will be inserted by publisher] for details of the index computation for the Weyl anomaly, and for the contributions of a range of massless fields to the gravitational anomaly on twistor space. The Supplemental Material includes Refs. [47-51]}\BibitemShut {NoStop}%
\bibitem [{\citenamefont {Costello}(2021)}]{Costello:2021bah}%
  \BibitemOpen
  \bibfield  {author} {\bibinfo {author} {\bibfnamefont {K.~J.}\ \bibnamefont {Costello}},\ }\href@noop {} {\enquote {\bibinfo {title} {{Quantizing Local Holomorphic Field Theories on Twistor Space}},}\ } (\bibinfo {year} {2021}),\ \bibinfo {note} {preprint},\ \Eprint {http://arxiv.org/abs/2111.08879} {arXiv:2111.08879 [hep-th]} \BibitemShut {NoStop}%
\bibitem [{\citenamefont {Caswell}(1974)}]{Caswell:1974gg}%
  \BibitemOpen
  \bibfield  {author} {\bibinfo {author} {\bibfnamefont {W.~E.}\ \bibnamefont {Caswell}},\ }\href {\doibase 10.1103/PhysRevLett.33.244} {\bibfield  {journal} {\bibinfo  {journal} {Phys. Rev. Lett.}\ }\textbf {\bibinfo {volume} {33}},\ \bibinfo {pages} {244} (\bibinfo {year} {1974})}\BibitemShut {NoStop}%
\bibitem [{\citenamefont {Jones}(1974)}]{Jones:1974mm}%
  \BibitemOpen
  \bibfield  {author} {\bibinfo {author} {\bibfnamefont {D.~R.~T.}\ \bibnamefont {Jones}},\ }\href {\doibase 10.1016/0550-3213(74)90093-5} {\bibfield  {journal} {\bibinfo  {journal} {Nucl. Phys. B}\ }\textbf {\bibinfo {volume} {75}},\ \bibinfo {pages} {531} (\bibinfo {year} {1974})}\BibitemShut {NoStop}%
\bibitem [{\citenamefont {Jones}(1975)}]{Jones:1974pg}%
  \BibitemOpen
  \bibfield  {author} {\bibinfo {author} {\bibfnamefont {D.~R.~T.}\ \bibnamefont {Jones}},\ }\href {\doibase 10.1016/0550-3213(75)90256-4} {\bibfield  {journal} {\bibinfo  {journal} {Nucl. Phys. B}\ }\textbf {\bibinfo {volume} {87}},\ \bibinfo {pages} {127} (\bibinfo {year} {1975})}\BibitemShut {NoStop}%
\bibitem [{Note3()}]{Note3}%
  \BibitemOpen
  \bibinfo {note} {The two-loop flow of the canonical coupling also receives contributions from Yukawa couplings which are difficult to incorporate on twistor space \cite {Jones:1981we,Machacek:1983tz}.}\BibitemShut {Stop}%
\bibitem [{\citenamefont {Novikov}\ \emph {et~al.}(1986)\citenamefont {Novikov}, \citenamefont {Shifman}, \citenamefont {Vainshtein},\ and\ \citenamefont {Zakharov}}]{Novikov:1985rd}%
  \BibitemOpen
  \bibfield  {author} {\bibinfo {author} {\bibfnamefont {V.~A.}\ \bibnamefont {Novikov}}, \bibinfo {author} {\bibfnamefont {M.~A.}\ \bibnamefont {Shifman}}, \bibinfo {author} {\bibfnamefont {A.~I.}\ \bibnamefont {Vainshtein}}, \ and\ \bibinfo {author} {\bibfnamefont {V.~I.}\ \bibnamefont {Zakharov}},\ }\href {\doibase 10.1016/0370-2693(86)90810-5} {\bibfield  {journal} {\bibinfo  {journal} {Phys. Lett. B}\ }\textbf {\bibinfo {volume} {166}},\ \bibinfo {pages} {329} (\bibinfo {year} {1986})}\BibitemShut {NoStop}%
\bibitem [{\citenamefont {Arkani-Hamed}\ and\ \citenamefont {Murayama}(2000)}]{Arkani-Hamed:1997qui}%
  \BibitemOpen
  \bibfield  {author} {\bibinfo {author} {\bibfnamefont {N.}~\bibnamefont {Arkani-Hamed}}\ and\ \bibinfo {author} {\bibfnamefont {H.}~\bibnamefont {Murayama}},\ }\href {\doibase 10.1088/1126-6708/2000/06/030} {\bibfield  {journal} {\bibinfo  {journal} {JHEP}\ }\textbf {\bibinfo {volume} {06}},\ \bibinfo {pages} {030} (\bibinfo {year} {2000})},\ \Eprint {http://arxiv.org/abs/hep-th/9707133} {arXiv:hep-th/9707133} \BibitemShut {NoStop}%
\bibitem [{\citenamefont {LeBrun}(1991)}]{lebrun1991explicit}%
  \BibitemOpen
  \bibfield  {author} {\bibinfo {author} {\bibfnamefont {C.}~\bibnamefont {LeBrun}},\ }\href@noop {} {\bibfield  {journal} {\bibinfo  {journal} {J. Differential Geometry}\ }\textbf {\bibinfo {volume} {34}},\ \bibinfo {pages} {223} (\bibinfo {year} {1991})}\BibitemShut {NoStop}%
\bibitem [{\citenamefont {Bittleston}\ \emph {et~al.}(2023)\citenamefont {Bittleston}, \citenamefont {Sharma},\ and\ \citenamefont {Skinner}}]{Bittleston:2022nfr}%
  \BibitemOpen
  \bibfield  {author} {\bibinfo {author} {\bibfnamefont {R.}~\bibnamefont {Bittleston}}, \bibinfo {author} {\bibfnamefont {A.}~\bibnamefont {Sharma}}, \ and\ \bibinfo {author} {\bibfnamefont {D.}~\bibnamefont {Skinner}},\ }\href {\doibase 10.1007/s00220-023-04828-0} {\bibfield  {journal} {\bibinfo  {journal} {Commun. Math. Phys.}\ }\textbf {\bibinfo {volume} {403}},\ \bibinfo {pages} {1543} (\bibinfo {year} {2023})},\ \Eprint {http://arxiv.org/abs/2208.12701} {arXiv:2208.12701 [hep-th]} \BibitemShut {NoStop}%
\bibitem [{Note4()}]{Note4}%
  \BibitemOpen
  \bibinfo {note} {Although we note that the twistorial anomaly coefficient $\chi $ is proportional to the one-loop vacuum energy \cite {Boyle:2021jaz}. It's not clear to us if $\psi $ also has a space-time avatar.}\BibitemShut {Stop}%
\bibitem [{\citenamefont {Riegert}(1984)}]{Riegert:1984kt}%
  \BibitemOpen
  \bibfield  {author} {\bibinfo {author} {\bibfnamefont {R.~J.}\ \bibnamefont {Riegert}},\ }\href {\doibase 10.1016/0370-2693(84)90983-3} {\bibfield  {journal} {\bibinfo  {journal} {Phys. Lett. B}\ }\textbf {\bibinfo {volume} {134}},\ \bibinfo {pages} {56} (\bibinfo {year} {1984})}\BibitemShut {NoStop}%
\bibitem [{\citenamefont {Jones}(1982)}]{Jones:1981we}%
  \BibitemOpen
  \bibfield  {author} {\bibinfo {author} {\bibfnamefont {D.~R.~T.}\ \bibnamefont {Jones}},\ }\href {\doibase 10.1103/PhysRevD.25.581} {\bibfield  {journal} {\bibinfo  {journal} {Phys. Rev. D}\ }\textbf {\bibinfo {volume} {25}},\ \bibinfo {pages} {581} (\bibinfo {year} {1982})}\BibitemShut {NoStop}%
\bibitem [{\citenamefont {Machacek}\ and\ \citenamefont {Vaughn}(1983)}]{Machacek:1983tz}%
  \BibitemOpen
  \bibfield  {author} {\bibinfo {author} {\bibfnamefont {M.~E.}\ \bibnamefont {Machacek}}\ and\ \bibinfo {author} {\bibfnamefont {M.~T.}\ \bibnamefont {Vaughn}},\ }\href {\doibase 10.1016/0550-3213(83)90610-7} {\bibfield  {journal} {\bibinfo  {journal} {Nucl. Phys. B}\ }\textbf {\bibinfo {volume} {222}},\ \bibinfo {pages} {83} (\bibinfo {year} {1983})}\BibitemShut {NoStop}%
\bibitem [{\citenamefont {Boyle}\ and\ \citenamefont {Turok}(2021)}]{Boyle:2021jaz}%
  \BibitemOpen
  \bibfield  {author} {\bibinfo {author} {\bibfnamefont {L.}~\bibnamefont {Boyle}}\ and\ \bibinfo {author} {\bibfnamefont {N.}~\bibnamefont {Turok}},\ }\href@noop {} {\enquote {\bibinfo {title} {{Cancelling the Vacuum Energy and Weyl Anomaly in the Standard Model with Dimension-Zero Scalar Fields}},}\ } (\bibinfo {year} {2021}),\ \bibinfo {note} {preprint},\ \Eprint {http://arxiv.org/abs/2110.06258} {arXiv:2110.06258 [hep-th]} \BibitemShut {NoStop}%
\bibitem [{\citenamefont {Berkovits}\ and\ \citenamefont {Witten}(2004)}]{Berkovits:2004jj}%
  \BibitemOpen
  \bibfield  {author} {\bibinfo {author} {\bibfnamefont {N.}~\bibnamefont {Berkovits}}\ and\ \bibinfo {author} {\bibfnamefont {E.}~\bibnamefont {Witten}},\ }\href {\doibase 10.1088/1126-6708/2004/08/009} {\bibfield  {journal} {\bibinfo  {journal} {JHEP}\ }\textbf {\bibinfo {volume} {08}},\ \bibinfo {pages} {009} (\bibinfo {year} {2004})},\ \Eprint {http://arxiv.org/abs/hep-th/0406051} {arXiv:hep-th/0406051} \BibitemShut {NoStop}%
\bibitem [{\citenamefont {Fradkin}\ and\ \citenamefont {Tseytlin}(1982{\natexlab{a}})}]{Fradkin:1981jc}%
  \BibitemOpen
  \bibfield  {author} {\bibinfo {author} {\bibfnamefont {E.~S.}\ \bibnamefont {Fradkin}}\ and\ \bibinfo {author} {\bibfnamefont {A.~A.}\ \bibnamefont {Tseytlin}},\ }\href {\doibase 10.1016/0550-3213(82)90481-3} {\bibfield  {journal} {\bibinfo  {journal} {Nucl. Phys. B}\ }\textbf {\bibinfo {volume} {203}},\ \bibinfo {pages} {157} (\bibinfo {year} {1982}{\natexlab{a}})}\BibitemShut {NoStop}%
\bibitem [{\citenamefont {Fradkin}\ and\ \citenamefont {Tseytlin}(1982{\natexlab{b}})}]{Fradkin:1982xc}%
  \BibitemOpen
  \bibfield  {author} {\bibinfo {author} {\bibfnamefont {E.~S.}\ \bibnamefont {Fradkin}}\ and\ \bibinfo {author} {\bibfnamefont {A.~A.}\ \bibnamefont {Tseytlin}},\ }\href {\doibase 10.1016/0370-2693(82)91018-8} {\bibfield  {journal} {\bibinfo  {journal} {Phys. Lett. B}\ }\textbf {\bibinfo {volume} {110}},\ \bibinfo {pages} {117} (\bibinfo {year} {1982}{\natexlab{b}})},\ \bibinfo {note} {[Erratum: Phys.Lett.B 126, (1983)]}\BibitemShut {NoStop}%
\bibitem [{\citenamefont {Fradkin}\ and\ \citenamefont {Tseytlin}(1984)}]{Fradkin:1983tg}%
  \BibitemOpen
  \bibfield  {author} {\bibinfo {author} {\bibfnamefont {E.~S.}\ \bibnamefont {Fradkin}}\ and\ \bibinfo {author} {\bibfnamefont {A.~A.}\ \bibnamefont {Tseytlin}},\ }\href {\doibase 10.1016/0370-2693(84)90668-3} {\bibfield  {journal} {\bibinfo  {journal} {Phys. Lett. B}\ }\textbf {\bibinfo {volume} {134}},\ \bibinfo {pages} {187} (\bibinfo {year} {1984})}\BibitemShut {NoStop}%
\bibitem [{\citenamefont {Witten}(2004)}]{Witten:2003nn}%
  \BibitemOpen
  \bibfield  {author} {\bibinfo {author} {\bibfnamefont {E.}~\bibnamefont {Witten}},\ }\href {\doibase 10.1007/s00220-004-1187-3} {\bibfield  {journal} {\bibinfo  {journal} {Commun. Math. Phys.}\ }\textbf {\bibinfo {volume} {252}},\ \bibinfo {pages} {189} (\bibinfo {year} {2004})},\ \Eprint {http://arxiv.org/abs/hep-th/0312171} {arXiv:hep-th/0312171} \BibitemShut {NoStop}%
\end{thebibliography}%
\bibliographystyle{apsrev4-1}
\clearpage


\widetext
\clearpage
\begin{center}
\textbf{\large Supplemental Materials: \\ Computing Weyl Anomalies}
\end{center}

We provide some details for the computation of the index \eqref{eqn:GH_index} and collect values of the constants $a,c,\chi,\psi$ appearing in the anomaly polynomial \eqref{eqn:sdcg_anomaly}. 

\setcounter{equation}{0}
\setcounter{figure}{0}
\setcounter{page}{1}
\setcounter{section}{0}
\setcounter{table}{0}
\makeatletter
\renewcommand{\theequation}{S\arabic{equation}}
\renewcommand{\thefigure}{S\arabic{figure}}


\section{Index Computation} \label{supp:GH_index}

Here we evaluate the index \eqref{eqn:GH_index}. The first step is computing the Todd class of $T_{\cPT_L/\cM^\text{GH}_L}$, which follows from the short exact sequence \eqref{eqn:additivity}. By additivity
\be \Todd\big(T_{\cPT_L}/\cM^\text{GH}_L\big) = \Todd(L\otimes\cN)\Todd(\Oo(2))\,. \ee
We're interested in contributions to the anomaly polynomial of the form $\chern_1(L)H\cha_2(\cN)$, and the only possible contribution to the index involving the Chern character of the normal bundle is from $\Todd(L\otimes\cN)$. Suppressing arguments for the moment, recall that
\be
\Todd = 1 + \frac{1}{2}\chern_1 + \frac{1}{12}(\chern_1^2+\chern_2) + \frac{1}{24}\chern_1\chern_2 - \frac{1}{720}(\chern_1^4 - 4\chern_1^2\chern_2 - \chern_1\chern_3 - 3\chern_2^2 + \chern_4) + \dots\,.
\ee
Since $L\otimes\cN$ has rank two both $\chern_3(L\otimes\cN)$ and $\chern_4(L\otimes\cN) = 0$. It's immediate that $\Todd(\Oo(2)) = 1 + H$ and
\be \chern_1(L\otimes\cN) = \chern_1(\cN) + 2\chern_1(L) = 2H + 2\chern_1(L)\,. \ee
Marginally less straightforwardly
\bea
&\chern_2(L\otimes\cN) = \frac{1}{2}\chern_1(L\otimes\cN)^2 - \cha_2(L\otimes\cN) \\
&= \frac{1}{2}(\chern_1(\cN) + 2\chern_1(L))^2 - \cha_2(\cN) - \chern_1(\cN)\chern_1(L) - 2\cha_2(L) = \chern_1(L)^2 + 2H\chern_1(L) - \cha_2(\cN)\,,
\eea
where we've used the fact that $H^2=0$. Retaining only terms in the Todd class that are at most first order in $\chern_1(L)$ and linear in $\cha_2(\cN)$ we have
\bea
&\Todd(L\otimes\cN) = - \frac{1}{12}\cha_2(\cN) - \frac{1}{12}(H + \chern_1(L))\cha_2(\cN) \\
&- \frac{1}{720}\big(16(H+\chern_1(L))^2 + 12H\chern_1(L)\big)\cha_2(\cN) = - \frac{1}{12}\bigg(1 + H + \chern_1(L) + \frac{11}{15}H\chern_1(L)\bigg)\cha_2(\cN)\,.
\eea
In full the relevant terms in the Todd class of the vertical tangent bundle are
\be \Todd\big(T_{\cPT_L}/\cM^\text{GH}_L\big) = \Todd(\cN\otimes L)\Todd(\Oo(2)) = - \frac{1}{12}\bigg(1 + 2H + \chern_1(L) + \frac{26}{15}H\chern_1(L)\bigg)\cha_2(\cN)\,. \ee
The Chern character in equation \eqref{eqn:GH_index} is not sensitive to $\cha_2(\cN)$, so we need only worry about terms involving at most one $H$ and $\chern_1(L)$. Bearing this in mind, we can compute the contributions of each field to half of the Chern character of $R$
\begin{itemize}
    \item The gauge boson bundle $\fsl(N_c)\otimes(\Oo\oplus K)$ contributes
    \be \op{rk}(\fsl(E))\big(1-2H-\chern_1(L) + 4\chern_1(L)H\big)\,. \ee
    \item The Dirac fermion bundle $\Pi(E\oplus E^\vee)\otimes(K^{1/4}\oplus K^{3/4})\otimes\C^{N_f}$ contributes
    \be  -2N_f\op{rk}(E)\Big(1-2H-\chern_1(L) + \frac{5}{2}\chern_1(L)H\Big)\,. \ee
    \item The complex scalars, contributing $R = (E\oplus E^\vee)\otimes K^{1/2}\otimes\C^{N_s}$:
    \be  N_s\op{rk}(E)\big(1-2H-\chern_1(L) + 2\chern_1(L)H\big)\,. \ee
\end{itemize}
These are similar to the contributions in a gauge instanton background, though with the second Chern character of the gauge bundle replaced by its rank. Putting these together yields equation \eqref{eqn:evaluate_index} in the main text.


\section{Twistorial Anomaly Coefficients} \label{supp:sdcg-anomaly}

Here we collect the contribution of various fields to the twistorial anomaly coefficients of self-dual conformal conformal gravity \eqref{eqn:sdcg_anomaly}. (All have been evaluated using index theory, detailed computations will appear elsewhere.) Those of conformally coupled scalars, Weyl fermions and gluons are listed in the table below.
\setlength\tabcolsep{7.5pt}
\begin{table}[h!]
    \centering
    \begin{tabular}{l c r r r r}
	\toprule
	Spin & Twistor Bundle & $a$ & $c$ & $\psi$& $\chi$ \\ \midrule
	$s=0$ & $(K^{1/2})^{\oplus2}$ & $\frac{1}{180}$ & $\frac{1}{60}$ & $0$ & $2$ \\
	$s=\frac{1}{2}$ & $\Pi(K^{1/4}\oplus K^{3/4})$ & $\frac{11}{720}$ & $\frac{1}{40}$ & $0$ & $-2$ \\
	$s=1$ & $\mscr{O}\oplus K$ & $\frac{31}{180}$ & $\frac{1}{10}$ & $0$ & $2$ \\
	\bottomrule
    \end{tabular}
    \caption{Anomaly Coefficients for Ordinary Fields}
    \label{tab:ordinary_coefficients}
\end{table}

We can also find the contributions of higher derivative fields appearing in the supermutliplet of $\cN=4$ self-dual superconformal gravity (whose twistor uplift was introduced in \cite{Berkovits:2004jj}). These include conformal gravitons, conformal gravitinos, antisymmetric tensors, higher derivative Weyl fermions and dimension zero complex scalars. Their contributions are listed in the table below. $\Omega^{2,0}_\p$ denotes the sheaf of closed $(2,0)$-forms.
\begin{table}[h!]
    \centering
    \begin{tabular}{l c r r r r r}
	\toprule
	Spin & Twistor Bundle & $a$ & $c$ & $\psi$ & $\chi$ \\ \midrule
	$s=0$ & $\Omega^{2,0}_\p$ & $-\frac{7}{45}$ & $-\frac{2}{15}$ & $-\frac{1}{12}$ & $4$ \\
	$s=\frac{1}{2}$ & $\Pi\big((T\otimes K^{3/4})\oplus(T^\vee\otimes K^{1/4})\big)$ & $-\frac{3}{80}$ & $-\frac{1}{120}$ & $-\frac{1}{24}$ & $-6$ \\
	$s=1$ & $(T\oplus T^\vee)\otimes K^{1/2}$ & $-\frac{19}{60}$ & $\frac{1}{20}$ & $\frac{1}{6}$ & $6$ \\
	$s=\frac{3}{2}$ & $\Pi\big((T\otimes K^{1/4})\oplus(T^\vee\otimes K^{3/4})\oplus K^{-1/4}\oplus K^{5/4}\big)$ & $-\frac{137}{90}$ & $-\frac{149}{60}$ & $-\frac{7}{24}$ & $-8$ \\
	$s=2$ & $T\oplus(T^\vee\otimes K)$ & $\frac{87}{20}$ & $\frac{199}{30}$ & $\frac{5}{12}$ & $6$ \\
	\bottomrule
    \end{tabular}
    \caption{Anomaly Coefficients for Higher Derivative Fields}
\label{tab:higher_derivative_coefficients}
\end{table}

The $a,c$ and $\chi$ coefficients appearing in tables \eqref{tab:ordinary_coefficients} and \eqref{tab:higher_derivative_coefficients} match those in the literature \cite{Fradkin:1981jc,Fradkin:1982xc,Fradkin:1983tg}. The $\psi$ coefficients are novel. We note that they sum to zero in the twistor uplift of $\cN=4$ self-dual superconformal gravity as necessitated by consistency of the twistor string \cite{Witten:2003nn,Berkovits:2004jj}.


\end{document}